\documentclass[twoside,twocolumn,9pt]{article}
\usepackage{extsizes}
\usepackage[super,sort&compress,comma]{natbib} 
\usepackage[version=3]{mhchem}
\usepackage[left=1.5cm, right=1.5cm, top=1.785cm, bottom=2.0cm]{geometry}
\usepackage{balance}
\usepackage{mathptmx}
\usepackage{amsfonts}
\usepackage{sectsty}
\usepackage{graphicx} 
\usepackage{lastpage}
\usepackage[format=plain,justification=justified,singlelinecheck=false,font={stretch=1.125,small,sf},labelfont=bf,labelsep=space]{caption}
\usepackage{float}
\usepackage{fancyhdr}
\usepackage{fnpos}
\usepackage[english]{babel}
\usepackage{algorithmic}
\usepackage{algorithm}
\addto{\captionsenglish}{%
  
}
\usepackage{array}
\usepackage{droidsans}
\usepackage{charter}
\usepackage[T1]{fontenc}
\usepackage[usenames,dvipsnames]{xcolor}
\usepackage{setspace}
\usepackage[compact]{titlesec}
\usepackage{hyperref}

\definecolor{cream}{RGB}{222,217,201}

\begin{document}

\pagestyle{fancy}
\thispagestyle{plain}
\fancypagestyle{plain}{
\renewcommand{\headrulewidth}{0pt}
}

\makeFNbottom
\makeatletter
\renewcommand\LARGE{\@setfontsize\LARGE{15pt}{17}}
\renewcommand\Large{\@setfontsize\Large{12pt}{14}}
\renewcommand\large{\@setfontsize\large{10pt}{12}}
\renewcommand\footnotesize{\@setfontsize\footnotesize{7pt}{10}}
\makeatother

\renewcommand{\thefootnote}{\fnsymbol{footnote}}
\renewcommand\footnoterule{\vspace*{1pt}%
\color{cream}\hrule width 3.5in height 0.4pt \color{black}\vspace*{5pt}} 
\setcounter{secnumdepth}{5}

\makeatletter 
\renewcommand\@biblabel[1]{#1}            
\renewcommand\@makefntext[1]%
{\noindent\makebox[0pt][r]{\@thefnmark\,}#1}
\makeatother 
\renewcommand{\figurename}{\small{Fig.}~}
\sectionfont{\sffamily\Large}
\subsectionfont{\normalsize}
\subsubsectionfont{\bf}
\setstretch{1.125} 
\setlength{\skip\footins}{0.8cm}
\setlength{\footnotesep}{0.25cm}
\setlength{\jot}{10pt}
\titlespacing*{\section}{0pt}{4pt}{4pt}
\titlespacing*{\subsection}{0pt}{15pt}{1pt}

\fancyfoot{}
\fancyfoot[RO]{\footnotesize{\sffamily{2023 ~\textbar  \hspace{2pt}\thepage}}}
\fancyhead{}
\renewcommand{\headrulewidth}{0pt} 
\renewcommand{\footrulewidth}{0pt}
\setlength{\arrayrulewidth}{1pt}
\setlength{\columnsep}{6.5mm}
\setlength\bibsep{1pt}

\makeatletter 
\newlength{\figrulesep} 
\setlength{\figrulesep}{0.5\textfloatsep} 

\newcommand{\topfigrule}{\vspace*{-1pt}%
\noindent{\color{cream}\rule[-\figrulesep]{\columnwidth}{1.5pt}} }

\newcommand{\botfigrule}{\vspace*{-2pt}%
\noindent{\color{cream}\rule[\figrulesep]{\columnwidth}{1.5pt}} }

\newcommand{\dblfigrule}{\vspace*{-1pt}%
\noindent{\color{cream}\rule[-\figrulesep]{\textwidth}{1.5pt}} }

\makeatother

\twocolumn[
  \begin{@twocolumnfalse}
\vspace{1em}
\sffamily
\begin{tabular}{p{18cm} }
\hline
\vspace{0.6cm} \\
\noindent\LARGE{\textbf{Measuring city-scale green infrastructure drawdown dynamics using internet-connected sensors in Detroit}} \\
\vspace{0.6cm} \\
\hline
\vspace{1cm} \\

 \noindent\large{Brooke E. Mason,$^{\ast}$\textit{$^{a}$} and Jacquelyn Schmidt\textit{$^{a}$} and Branko Kerkez\textit{$^{a}$}} \\

\vspace{0.6cm} \\

\noindent\normalsize{
The impact of green infrastructure (GI) on the urban drainage landscape remains largely unmeasured at high temporal and spatial scales. To that end, a data toolchain is introduced, underpinned by a novel wireless sensor network for continuously measuring real-time water levels in GI. The internet-connected sensors enable the collection of high-resolution data across large regions. A case study in Detroit (MI, US) is presented, where the water levels of 14 GI sites were measured in-situ from June to September 2021. The large dataset is analyzed using an automated storm segmentation methodology, which automatically extracts and analyzes individual storms from measurement time series.  Storms are used to parameterize a dynamical system model of GI drawdown dynamics. The model is completely described by the decay constant $\alpha$, which is directly proportional to the drawdown rate. The parameter is analyzed across storms to compare GI dynamics between sites and to determine the major design and physiographic features that drive drawdown dynamics. A correlation analysis using Spearman’s rank correlation coefficient reveals that depth to groundwater, imperviousness, longitude, and drainage area to surface area ratio are the  most important features explaining GI drawdown dynamics in Detroit. A discussion is provided to contextualize these finding and explore the implications of data-driven strategies for GI design and placement.}

\end{tabular}

 \end{@twocolumnfalse} \vspace{0.6cm}

  ]

\renewcommand*\rmdefault{bch}\normalfont\upshape
\rmfamily
\section*{}
\vspace{-1cm}


\footnotetext{\textit{$^{a}$University of Michigan, Department of Civil and Environmental Engineering, 2350 Hayward St, Ann Arbor, Michigan 48109, US; E-mail: bemason@umich.edu}} 



\section{Water Impact Statement}
Globally, green infrastructure (GI) has become a popular stormwater management solution, but its impact on the larger urban drainage landscape remains unverified. A low-cost, low-maintenance sensor is introduced for real-time, high-resolution GI monitoring. When coupled with an automated data toolchain, we show how investments in monitoring networks support a more targeted and data-driven approach to GI placement, planning, and maintenance. 

\section{Introduction}
Urban areas around the world are struggling to manage stormwater runoff and flooding-- a challenge compounded by rapid urbanization and climate change. \cite{cohen_urbanization_2006, zhu_climate_2007} Gray infrastructure, which consists of gutters, drains, and pipes, is the traditional method for collecting and conveying stormwater away from urban areas. Recently, green infrastructure (GI) has become a popular alternative, used either as a standalone stormwater management practice or in concert with traditional gray infrastructure. \cite{eckart_performance_2017, golden_green_2018} GI attempts to mimic the natural water cycle by using plants, soil, and landscape design to capture and filter local runoff.\cite{ahiablame_effectiveness_2012, eckart_performance_2017} One of the most common GI practices is bioretention cells, or rain gardens, which are depressed vegetated areas that capture and reduce runoff by allowing it to evapotranspire or exfiltrate into surrounding soil.\cite{hunt_meeting_2012} 

Communities worldwide are investing in GI for managing stormwater at increasing scales. For example, China plans to spend over US\$ 1.5 trillion on GI in 657 cities by 2030.\cite{jia_chinas_2017} In the midwestern US, the city of Detroit, Michigan invested US\$ 15 million in GI between 2013--2017 and will invest US\$ 50 million by 2029.\cite{detroit_GIprojects_2022} These investments assume adding more GI assets will positively impact stormwater outcomes, however, sufficient data to support this claim has yet to be produced.\cite{clary_integration_2011, ahiablame_effectiveness_2012, fletcher_understanding_2013, eckart_performance_2017}

Real-time monitoring of stormwater infrastructure at high temporal and spatial resolutions is now possible with Internet of Things (IoT) technologies.\cite{blumensaat_how_2019, berglund_smart_2020} Real-time sensing has been successfully deployed to monitor depths and flows in stormwater\cite{bartos_open_2018} and sewer networks.\cite{cembrano_optimal_2004, sun_real-time_2020} Recently, some studies have used sensors, such as pressure transducers connected to data loggers, to monitor GI.\cite{kazemi_assessment_2017, lewellyn_evaluation_2016, winston_quantifying_2016, zukowski_evaluation_2016} While these studies provided high resolution measurements, they required frequent field maintenance (e.g., downloading the data onsite, replacing batteries), making this approach impractical for obtaining large-scale, and/or long-term data. Therefore, there is still a need for GI IoT solutions.

To that end, we introduce an end-to-end data toolchain based on new wireless sensors for estimating real-time drawdown in GI, the speed at which stormwater is evapotranspired and exfiltrated into the native soil.\cite{winston_quantifying_2016, ahiablame_effectiveness_2012} These wireless sensors are low-cost, easy to install, and can be deployed at scale to create large, long-term, high-resolution datasets of urban drainage conditions. When combined with an analytics toolchain, our approach can be used to automatically learn GI dynamics from data on a storm-by-storm basis. To study the value of a city-wide dataset, we present a case study of these GI sensors deployed in Detroit. This novel dataset is used to characterize the drawdown dynamics of GI over multiple storms. The core contribution of this paper is a new sensor and data analysis methodology, along with experimental results that show which factors are the strongest predictors of drawdown dynamics for the studied GI network.

\section{Background}
\subsection{GI design standards}
Many communities rely on established stormwater management manuals, which detail how to select, design, construct, and maintain stormwater infrastructure, including GI. A manual’s goal is to set forth best management practices which will elicit a certain level of performance, such as mitigating peak flow or infiltrating a certain fraction of runoff.\cite{roy_impediments_2008} Regional and local manuals set design requirements (e.g., site selection, GI selection/sizing, soil media composition, underdrain sizing, plant selection) as well as performance metrics.\cite{hunt_meeting_2012} These design requirements and performance metrics exist for a variety of reasons, for example to ensure public safety and limit liability by eliminating trip hazards, adding barriers around water features, and reducing standing water to control mosquitos, but most fundamentally, to ensure that stormwater is being managed consistently across various sites. As an example, in the US, two common metrics for rain gardens and bioretention cells include the maximum allowable ponding time, generally 12--48 hours,\cite{mathews_post-construction_2021, new_york_city_environmental_protection_new_2022, water_and_sewerage_department_stormwater_2022} and infiltration rate, typically 2.5--5 cm/hr.\cite{hunt_meeting_2012, mathews_post-construction_2021, water_and_sewerage_department_stormwater_2022}

While infiltration rates can vary substantially even within the same GI, drawdown rates are representative of the entire system.\cite{ahmed_modified_2014, asleson_performance_2009} The drawdown rate of GI is a function of the design features, building and maintenance practices, and the surrounding and underlying physiographic features.\cite{ebrahimian_temporal_2020, eckart_performance_2017} Design features include size, soil type, and vegetation. During site construction, how the sites are excavated and graded can cause significant soil compaction which ultimately impacts GI drawdown rates.\cite{brown_impacts_2010} Physiographic features include the native soils, topography, land use type, depth to groundwater, and sunlight.\cite{us_environmental_protection_agency_reducing_2007, eckart_performance_2017} While GI design can be optimized, in most cases the surrounding physiographic features cannot be changed. These features may have a strong effect on GI drawdown. For example, a shallow groundwater table (< 2--3 m) may result in more saturated media, which forms a smaller hydraulic gradient, impeding infiltration into the GI and exfiltration out of the GI into surrounding native soil.\cite{jackisch_hydrologic_2017, zhang_evaluating_2017} This suggests that the drawdown rate of GI is governed by a complex interaction between its design features and surrounding physiographic features. Few large-scale data sets exist to verify these patterns at scale, however.

\subsection{GI measurements}
Monitoring is needed to confirm whether a GI is meeting desired management goals. Additionally, monitoring can be used to determine whether local stormwater manuals are setting appropriate design standards and performance metrics. Due to the sheer number of sites and the cost of measuring quantitative metrics such as drawdown rate, cities often rely on visual inspection or modeling to assess performance.\cite{ahiablame_effectiveness_2012} If GI monitoring is carried out, it is generally limited to certain time periods and conditions.\cite{ahiablame_effectiveness_2012, eckart_performance_2017, mitchell_applying_2006}

Drawdown rate has been traditionally measured via drawdown testing. A GI is filled with water (either synthetically or via rainfall) until ponding occurs, then the drain depth and time are recorded to calculate the drawdown rate.\cite{ebrahimian_temporal_2020, zukowski_evaluation_2016} These measurements are typically conducted manually with the help of a watch and gauge plate. Drawdown testing is generally only done pre- and post-installation,\cite{winston_quantifying_2016} but occasionally assets are tested as they age to track how they change over time.\cite{kluge_metal_2018, lewellyn_evaluation_2016} Unfortunately, the laboriousness of drawdown testing results in most communities having sparse datasets of in-situ GI drawdown. Furthermore, drawdown is inherently non-linear\cite{winston_quantifying_2016}, meaning that drawdown rate may change over the course of a storm and in response to ambient conditions. To gain a complete picture of GI behavior, more data are needed than what can be obtained from a single drawdown test taken during a single storm event.

Recent technological advances have opened up new possibilities for low-cost, high resolution stormwater sensing.\cite{berglund_smart_2020, blumensaat_how_2019} Despite their availability, the uptake of these technologies for GI management has been limited. According to a national survey of officials in water utilities and agencies, however, assumed high construction and maintenance costs associated with smart GI are the two main barriers to adoption.\cite{meng_stated_2019} As such, the concept has yet to be vetted at scale. 

\begin{figure*}[!htb]
\centering
  \includegraphics[width=\textwidth]{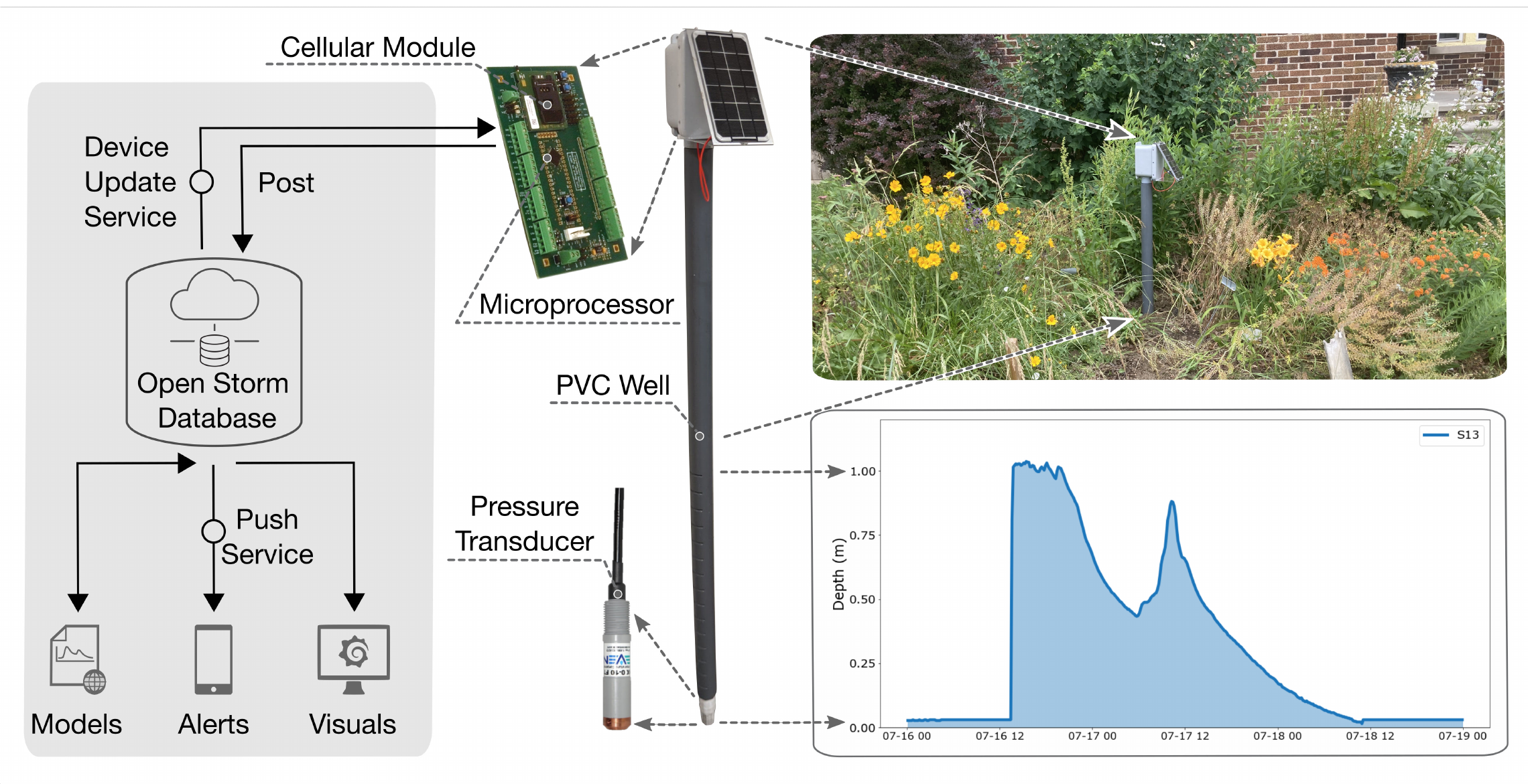}
  \caption{A GI sensor installed in a rain garden (top right). The sensor’s hardware layer (center) includes the PVC well, microcontroller, cellular modem, and pressure transducer. The cloud services layer (left) includes the database backend, along with applications for controlling sensor behavior and visualizing data (bottom right).}
  \label{fig:GInode}
\end{figure*}

\section{Materials and methods}
\subsection{Green infrastructure wireless sensors}
A wireless sensor was designed to continuously measure drawdown in GI (Fig. \ref{fig:GInode}). Specifically, the device measures water level fluctuations in real-time. At the time of writing, the sensor costs approximately US\$ 1,000 to build and US\$ 25 annually for telecommunication and data storage services. The form factor of the sensor is similar to a water well, consisting of a 1.5 m long, slotted PVC pipe with one end holding the sensor and the other holding the remaining hardware components. The sensor uses the vetted Open Storm hardware and cloud services stack detailed in Bartos et al. (2018).\cite{bartos_open_2018} The hardware layer relies on an ultra-low power ARM Cortex-M3 microcontroller (Cypress PSoC). The microcontroller manages the sensing and data transmission logic of the embedded system. The sensor measures water levels to a reported accuracy of ±0.762 cm using a pressure transducer (Stevens SDX 93720-110), which converts a barometric reading to a 4--20 milliampere (mA) output. The sensor is equalized for atmospheric pressure changes and was calibrated in the laboratory using a standard water column. The device is connected to the internet with a 4G LTE CAT-4 cellular modem (Nimbelink NL-SW-LTE). The cellular modem enables bi-directional communication between the sensor and a remote cloud-hosted web server. The device is powered using a 3.7 V lithium-ion battery (Tenergy) that is recharged by a solar panel (Adafruit 500). Power consumption measurements were used to confirm that when the device is on, power consumption is in the milli-amperage range and when the device is in sleep mode, it is in the micro-amperage range. With these power consumption numbers the sensor can stay in the field for up to 10 years without needing a battery replacement.

The main reason for field maintenance occurs if a sensor malfunctions. The first type of sensor malfunction is sensor drift, which is defined as a small temporal variation in the sensor output under unchanging conditions. Sensor drift can be detected in this case when the sensor’s “zero” reading changes over time. The other type of sensor malfunction occurs if a sensor provides a zero reading during periods of rainfall. There are several possible explanations for this malfunction. First, since the sensor operates by converting current to depth, there could be an issue with the analog circuitry resulting in inaccurate current measurements. Second, the sensor could be physically damaged during node assembly or deployment. Third, the sensor provides a venting tube for equalizing atmospheric pressure changes. Although a cap is added to the tube to keep moisture out, if the cap is faulty, condensation can enter the tube and cause inaccurate readings. Finally, the PVC well may clog with sediment. To rectify any of the above sensor malfunctions, the sensor is swapped for a new one, which only takes a few minutes of field work.

The sensor measurements were validated in the field using a gauge plate and digital, time-lapse photography by an outside consultant.\cite{dierks_developing_2019} During rain events, photos were taken of the ponded water and gauge plate measurement every ten minutes (ESI Fig. A1). There was an average alignment of 11 mm between the camera-recorded and sensor-recorded depth measurements (ESI Fig. A2).

Installation of the sensor takes less than 30 minutes by one person and requires digging a 1 m deep hole using a simple, off-the-shelf, handheld post hole digger. The sensor is placed in the hole and backfilled with soil. Real-time data begins streaming to a web dashboard as soon as the unit is deployed. The sensor is deployed such that an water level of 0 m indicates dry conditions, while a measurement above 1 m indicates water is ponding on the surface.

The sensor takes measurements every ten minutes and reports data to the server once every hour. Measurements are transmitted over the cellular network via a secure connection to a cloud-hosted server. Data and metadata are stored in an InfluxDB database.\cite{influxdata_inc_influxdb_2022} Measurements are then made available for visualization and sharing with partners through Grafana,\cite{grafana_labs_grafana_2022} a dashboarding software used to plot measured water level over time. Both InfluxDB and Grafana instances are hosted on an Amazon Web Services (AWS) Elastic Cloud Computing (EC2) instance.\cite{amazon_web_services_elastic_2022} The system is entirely open source and the complete codebase, hardware schematics, and how-to guides have been made available as part of this paper on \url{github.com/kLabUM/GI_Sensor_Node}.

\subsection{Automatically learning GI dynamics from data}
To enable comparisons between sites without losing temporal information due to averaging, we synthesize and parameterize a drawdown model automatically from data. We assume that water levels inside GI can be approximated as a first-order linear dynamical system, which evolves according to the differential equation:
\begin{equation}
\label{eqn:diff_eqn}
    \frac{dh}{dt} = \alpha h \; ; \; \alpha < 0
\end{equation}
where $h$ represents the water level in GI and $\alpha$ is the decay constant-- a measure of how fast the water level inside a GI recedes following a storm. In this formulation, this decay constant is directly proportional to drawdown rate and provides a single parameter that can be compared between sites. A relatively larger magnitude $\alpha$ corresponds to a faster rate of drawdown, while a smaller magnitude $\alpha$ corresponds to more slowly changing water levels. More relevant to cross comparisons between sites, however, is that $\alpha$ embeds both temporal and magnitude information in one parameter. In other words, two sites could have similar bulk performance metrics, such as average volume capture over 24 hours, but exhibit vastly different drawdown curves. As such, studying the decay constant $\alpha$ allows us to compare sites while taking advantage of the temporal granularity of our sensor data.

\begin{figure*}[!htb]
\centering
  \includegraphics[width=\textwidth]{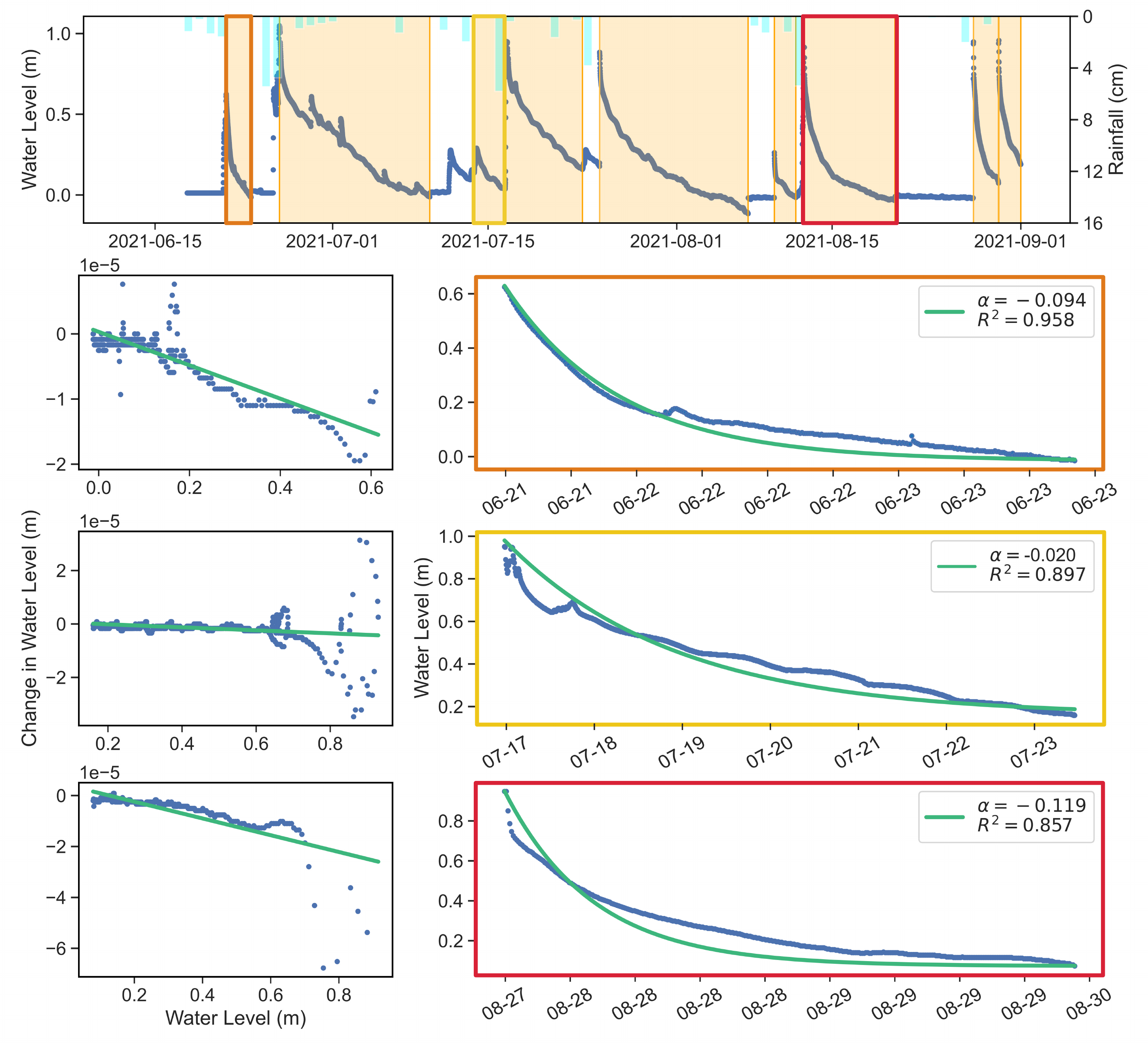}
  \caption{(Top row) Time series water level measurement from a GI overlaid with nearby publicly-available precipitation data. The orange boxes indicate distinct storm events automatically detected by a peak finding algorithm. The decay constant $\alpha$ is fit for three distinct storms in the same GI. (rows 2--4, left) To find $\alpha$, we fit a line for the relationship between water level (x-axis) and the change in water level (y-axis). (rows 2--4, right) The found $\alpha$'s are then plotted against the actual water levels experienced from the three distinct storms. The $R^{2}$ value for each fit is also provided.}
  \label{fig:fitting_decays}
\end{figure*}

Linear regression is used to fit the drawdown model to the water level sensor data of each storm. To fit the data to Eqn. \ref{eqn:diff_eqn}, we find the fit that best captures the relationship between the water level and its first derivative $[h(t),\frac{dh}{dt}]$ (Fig. \ref{fig:fitting_decays}, left col.). The slope of this line is the decay constant, $\alpha$. This method selects the most dominant rate of decay in the data. The fit of the model is evaluated using two metrics: the coefficient of determination (R$^{2}$) and root mean squared error (RMSE). To illustrate the methodology, the fit of the drawdown model to the sensor data for three distinct storms is shown in Fig. \ref{fig:fitting_decays}.

Since we calculate $\alpha$ for every storm, drawdown dynamics of each site can be compared on a storm-by-storm basis, or the set of $\alpha$’s can be combined into a single value for a given site. A single value of $\alpha$ can be thought of as a regression in $[h(t),\frac{dh}{dt}]$ feature space across all storms. This allows us to model the expected water level drawdown curve for a future storm. The resulting model could be used to inform estimates on how long a GI would take to drain given an initial water level of $h(0)$ m, for example. A parameterized decay model can also be used to simulate the GI’s behavior as part of a broader hydrologic simulator (e.g., US EPA SWMM\cite{rossman_storm_2015}).

\begin{figure*}[!htb]
\centering
  \includegraphics[width=\textwidth]{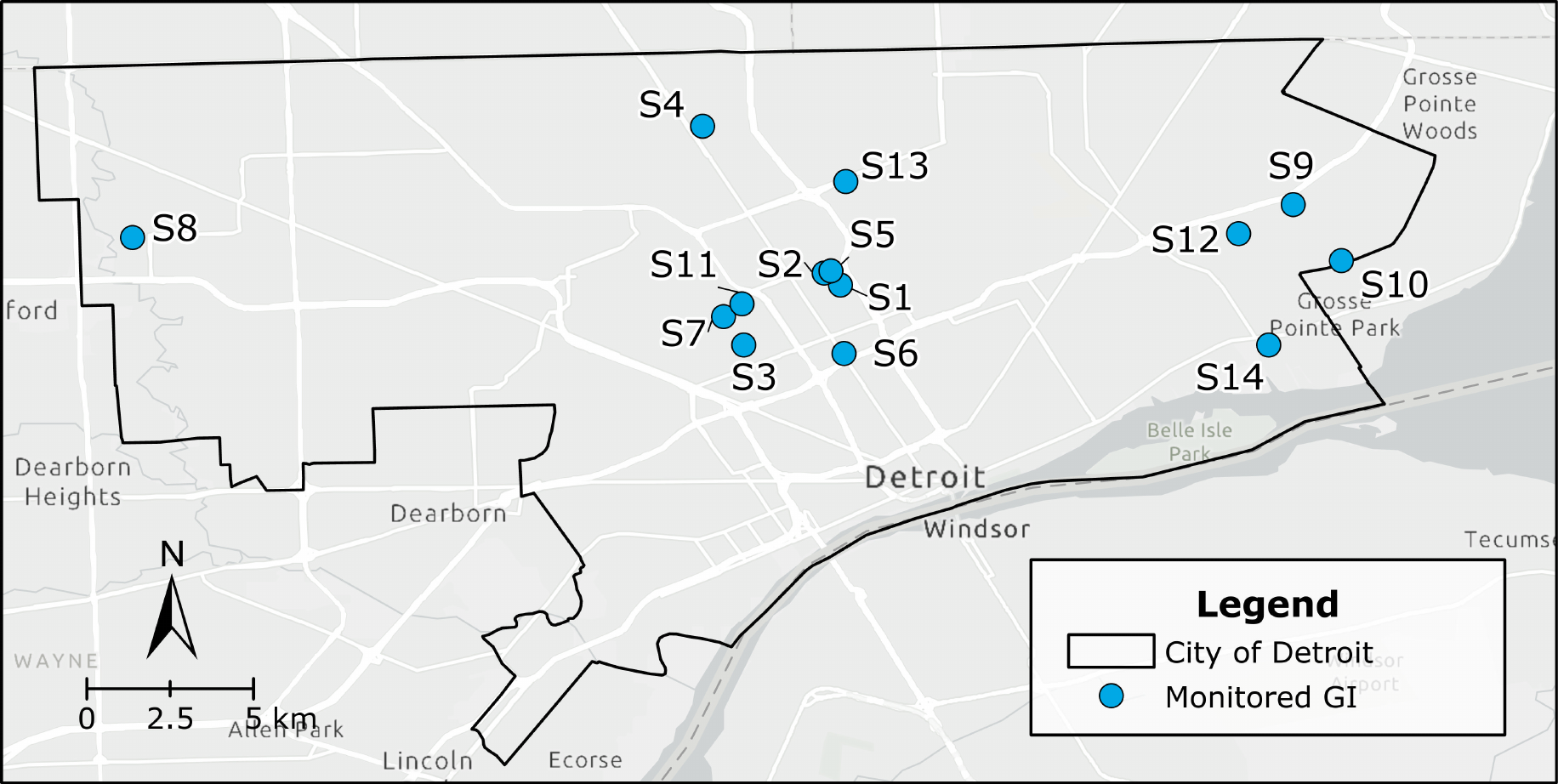}
  \caption{Map of the 14 GI sites selected for sensors in Detroit.}
  \label{fig:Detroit}
\end{figure*}

\subsubsection{Implementation}
An automated process is developed to identify individual storms in the sensor data. This methodology requires water level time data, in this case provided by our sensors. Storm events are automatically identified by marking local minima and maxima using the \texttt{find\textunderscore peaks()} function of Python Scipy Signal library.\cite{virtanen_scipy_2020} To find the maxima we pass the water level time series to the function, which returns a list of indices corresponding to peaks (local maxima). To find the minima, we pass the negative of the water level time series, which then returns a list of indicies for local minima. We use two of the function's optional parameters to refine which points qualify as "peaks": prominence ($p$) and distance. Prominence is a measure of how high a local maxima stands out in comparison to its neighboring local minima. The prominence parameter was adjusted for each site such that the selected peaks corresponded reasonably well to local rainfall measurements and captured a meaningful segment of water level drawdown for each storm. We set the distance parameter to 3 hours, meaning adjacent local minima/maxima must be at least 3 hours apart to be selected. An example of the resultant automated storm segmentation is provided in Fig. \ref{fig:fitting_decays}, top row. While rainfall data are not required for the method, they can nonetheless be used as a secondary check, by visually lining up storms detected in the water levels with those measured by nearby rain gages. 

Once the storms were isolated, the drawdown model is fit to the data using the \texttt{poly\textunderscore fit()} function of Python's Numpy library.\cite{harris_array_2020} The function uses least squares to fit a polynomial to the provided data. We pass $[h(t),\frac{dh}{dt}]$ to the function with the degree set to one. The function returns the $\alpha$ that minimizes the squared error. Fig. \ref{fig:fitting_decays} (rows 2--4) show these fits along with the resultant drawdown model for three storms measured at the same site. Taken out of the differential form, the drawdown model follows $x = Ce^{\alpha t} + b$, where $C$ and $b$ are scaling and offset parameters that are adjusted to fit the magnitude of the storm. The coefficient of determination (R$^{2}$) and root mean squared error (RMSE) are calculated for each fit using Python's Scikit-learn library.\cite{pedregosa_scikit-learn_2011} 

\subsection{Case study}
We selected Detroit, Michigan, US for the GI monitoring network (latitude 42°19'53'', longitude \textminus83°2'44''). Detroit has a unique opportunity for extensive GI installations because approximately 103 km$^{2}$ (28\%) of the city is classified as vacant land.\cite{meerow_spatial_2017} The city is located at the outlet of three major watersheds (i.e., Rouge River, Clinton River, Lake St. Clair) where flows eventually discharge into either Lake St. Clair or the Detroit River. Due to Detroit’s location in the floodplain, most of its soil is poorly drained clay and silt.\cite{mcfarland_guide_2019} Detroit also has a shallow groundwater table. Teimoori et al. (2021) found that the modeled depth to groundwater in Detroit ranged from approximately 1--3 meters below the ground surface.\cite{teimoori_modeling_2021} Detroit’s climate follows a four-season pattern, with average temperatures ranging from \textminus7.11°C to 28.7°C. Detroit averages 87 cm and 137 days of precipitation per year.\cite{national_weather_service_nowdata_2022} Precipitation is dispersed relatively evenly throughout the year as rain and snow, but heavier amounts occur in spring and winter.\cite{mcfarland_guide_2019}

Detroit has a combined sewer system for managing stormwater and wastewater which flows into the second largest wastewater plant in the world.\cite{mcfarland_guide_2019} During extreme rainfall events in 2021, the sewer conveyance and wastewater plant’s treatment capacity was exceeded on multiple occasions, resulting in billions of gallons of raw sewage being directly discharged into Detroit waterways.\cite{stein_150_2021} In addition, residential basements were flooded with sewage-laden runoff.\cite{stein_150_2021} The need to mitigate flooding and sewer overflows has driven the City of Detroit and organizations like the Detroit Sierra Club to prioritize GI installations.\cite{detroit_GIprojects_2022}

In partnership with the Detroit Sierra Club, a non-profit organization, 14 GI sites were selected for deployment in summer 2021 across 155 km$^{2}$ of Detroit to monitor GI performance (Fig. \ref{fig:Detroit}). Since 2015, the Detroit Sierra Club has been working with community partners and Detroit residents to build GI, primarily small residential rain gardens. GI were selected that varied in terms of age, size, and surrounding land use type. Twelve sites were rain gardens designed and built by Detroit Sierra Club and their partners, and two were engineered and commercially built bioretention cells. The design and site data for the GI were provided by Detroit Sierra Club (ESI Table A1). Moving forward, each site is identified by an alpha numeric code (e.g., S1 for site 1).

\subsection{Correlation analysis}
Once the decay constants were extracted from the Detroit sensor network, a correlation analysis was conducted to determine which design and physiographic features explain GI drawdown, as quantified by the decay constant $\alpha$. Design features included the GI’s location, surface area, drainage area, storage volume, soil media depth, age, and drainage area to surface area ratio (DA/SA ratio). The DA/SA ratio was calculated by dividing the drainage area by the surface area. The physiographic features for each GI were extracted from public GIS datasets of percent imperviousness, land use type, elevation, slope, native soil type (i.e., hydrologic soil group), and depth to groundwater. ESI Section B provides detailed steps on how the GIS datasets were downloaded, processed, and the features were extracted for each GI.

The datasets investigated included both non-normal continuous (e.g., surface area, elevation) and ordinal (e.g., land use type, hydrologic soil group) variables. To handle both types of variables, Spearman’s rank correlation coefficient was selected for the correlation analysis.\cite{forthofer_descriptive_2007} Spearman’s rank correlation coefficient is a nonparametric measure of the strength and direction of the monotonic relationship between two ranked variables.\cite{zar_spearman_2014}

Spearman's rank correlation coefficients were computed using the \texttt{corr()} function of Python's Pandas library.\cite{reback2020pandas} A dataframe of the mean decay constants, physiographic features, and design features for the GI monitoring network was passed to the function. The function requires a correlation method, which was set to \texttt{'spearman'}. Readers are directed to a Zenodo web portal to freely obtain the data and code referenced in this paper.\cite{mason_GI_2022}

\section{Results}

\subsection{Sensor network performance}
\label{Results:sensor_performance}
Deployment of the GI monitoring network began mid-June 2021 and 14 operational sensors were deployed by early July 2021 (installation dates provided in ESI Table A2). The measurement period consists of data collected between June 15, 2021, and September 1, 2021. During the measurement period, there were only two instances of prolonged data loss— S8 and S12 had a two-hour and 24-hour data gap, respectively. These losses did not impact the measurement of storm response at either site. Sensor drift was not an issue, with an average drift of < 2.5 cm. There was one maintenance trip on August 11th to swap S12’s sensor because it indicated the GI was empty during periods of rain (ESI Table A2).

\begin{figure*}[!htb]
\centering
  \includegraphics[width=\textwidth]{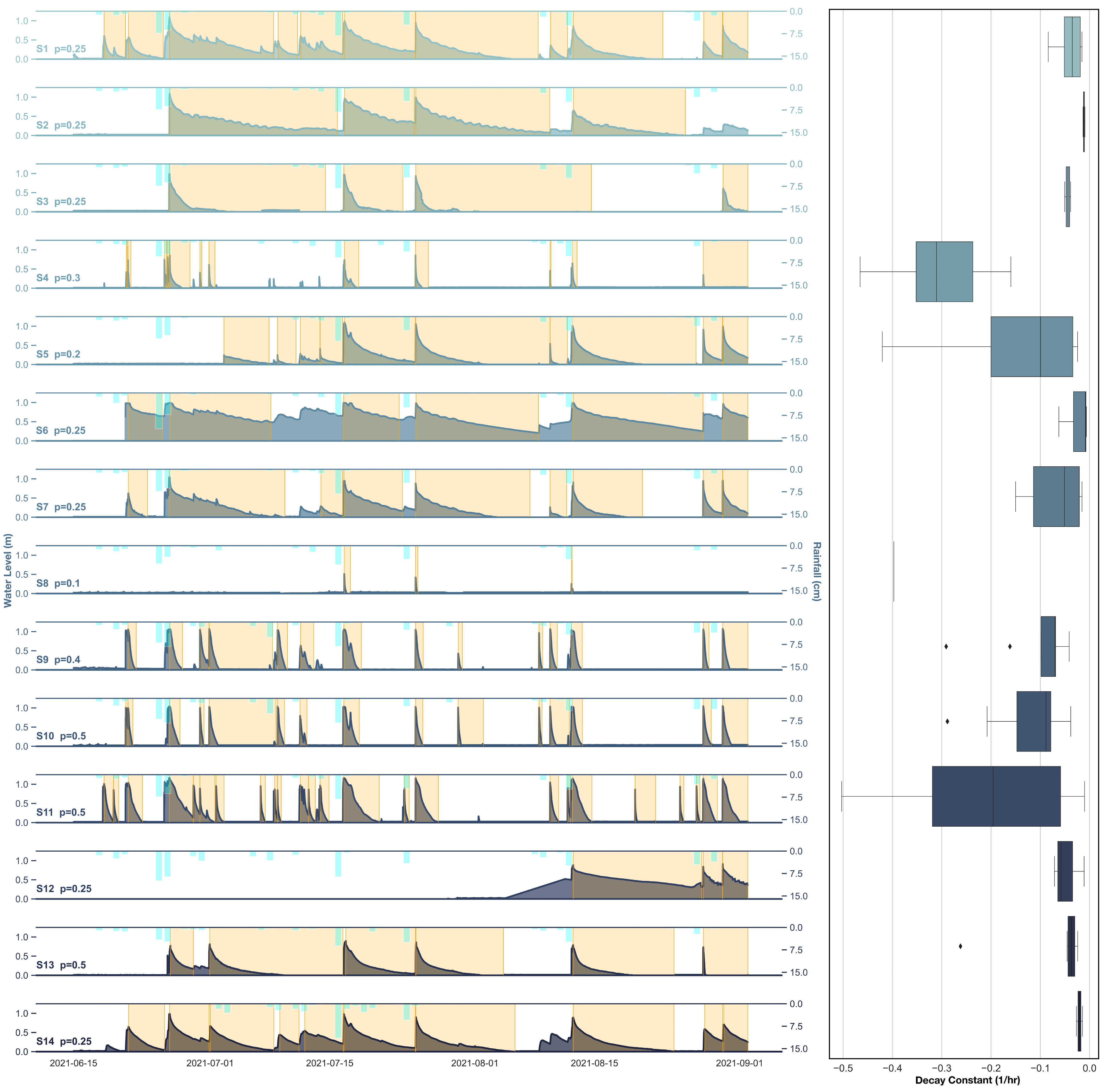}
  \caption{(left) Water level (m) measured across all sites on the left y-axis with rainfall (cm) on the right y-axis. Storm events are highlighted by the orange boxes. Prominence (p), the minimum increase in water level needed for a storm event to be considered distinct, is labeled for each site. (right) A boxplot showing the variance in each GI’s decay constants measured for all highlighted storms.}
  \label{fig:waterlevels}
\end{figure*}

\subsection{GI drawdown analysis}
\label{Results:drawdown}
The measurement period coincided with Detroit’s 7th wettest summer on record, which included several historic rain events: 15.2 cm of rain on June 25th, 5.6 cm on July 16th, and 6.9 cm on August 12th.\cite{hicks_summer_2021} During the measurement period, a total of 122 storms were identified across the network (orange boxes in Fig. \ref{fig:waterlevels} (left)). Of the 122 storms, 15 storms were excluded as outliers from the analysis due to poor fit of the drawdown model (negative R$^{2}$). A mean of 7.4 storm events were analyzed for each site with the number of distinct storm events varying widely per site: 21 for S11 versus 1 for S8. The variation in the number of storms captured by site is due to both the installation date (see ESI Table A2) and the spatial variation in rainfall.\cite{cristiano_spatial_2017}

\begin{table}[!htb]
    \centering
    \begin{tabular*}{0.48\textwidth}{@{\extracolsep{\fill}}ccccc}
    \hline
    Site & No. Storms & $\alpha$ & RMSE & $R^{2}$ \\
     & Analyzed & (mean) & (mean) & (mean) \\
    \hline
    S1	& 11/11 & \textminus0.040	&	5.159	& 0.834 \\
    S2	& 3/3 & \textminus0.011	&	6.306	& 0.875 \\
    S3	& 4/4 &  \textminus0.044	&	4.776	& 0.885 \\
    S4	& 9/12 & \textminus0.305	&	9.109	& 0.524 \\
    S5	& 9/9 & \textminus0.146	&	4.611	& 0.727 \\
    S6	& 5/6 & \textminus0.024	&	3.420	& 0.916 \\
    S7	& 9/9 & \textminus0.069	&	6.088	& 0.802 \\
    S8	& 1/3 & \textminus0.397	&	2.998	& 0.922 \\
    S9	& 9/12 & \textminus0.102	&	15.964	& 0.606 \\
    S10	& 11/12 & \textminus0.119	&	13.744	& 0.697 \\
    S11	& 21/24 & \textminus0.200	&	12.209	& 0.738 \\
    S12	& 3/3 & \textminus0.047	&	4.531	& 0.806 \\
    S13	& 6/6 & \textminus0.072	&	3.777	& 0.921 \\
    S14	& 7/8 & \textminus0.021	&	6.630	& 0.637 \\
    \hline
    \end{tabular*}
    \caption{The results from fitting the decay model for the storms captured by the GI monitoring network. We report the mean decay constant $\alpha$ for each GI and how well the decay constant $\alpha$ fit the sensor data as measured by RMSE and $R^{2}$.}
    \label{tab:model_fit}
\end{table}

The mean fit of the drawdown model to the sensor data was R$^{2}=$ 0.746 $\pm $0.111 and RMSE = 8.579 $\pm$ 4.168. The fitted decay constant $\alpha$ varied by storm and by GI (Fig. \ref{fig:waterlevels} (right)). Across all storms and sites, the mean decay constant $\alpha$ and standard deviation was \textminus0.119 $\pm$ 0.124 hr$^{-1}$. The average decay constant per site varied by two orders of magnitude, from \textminus0.011 hr$^{-1}$ (S2) to \textminus0.397 hr$^{-1}$ (S8). The number of storms identified versus analyzed, as well as the mean decay constant $\alpha$, RMSE, and R$^{2}$ for each GI is provided in Table \ref{tab:model_fit}.

The decay constant $\alpha$ corresponds with the GI's drainage dynamics. During the measurement period, most GI completely drained between storm events (S4, S8--S11), providing full storage for the next storm event (Fig. \ref{fig:waterlevels} (left)). S2, S6, and S12 always had some water present in their soil media, limiting the amount of storage for each subsequent storm. During the measurement period, most sites experienced ponding (water level > 1 m). However, ponding did not exceed 12 hours for most sites (11 of 14 sites). S6, S11, and S9 experienced extended periods of ponding during the June 25th storm for 22, 29, and 21 hours, respectively. Sites S6 and S11 also experienced extended ponding for approximately 24 hours during the July 16th storm, and S11 ponded for about 16 hours during the August 12th storm. 

\begin{figure*}[!htb]
\centering
  \includegraphics[width=\textwidth]{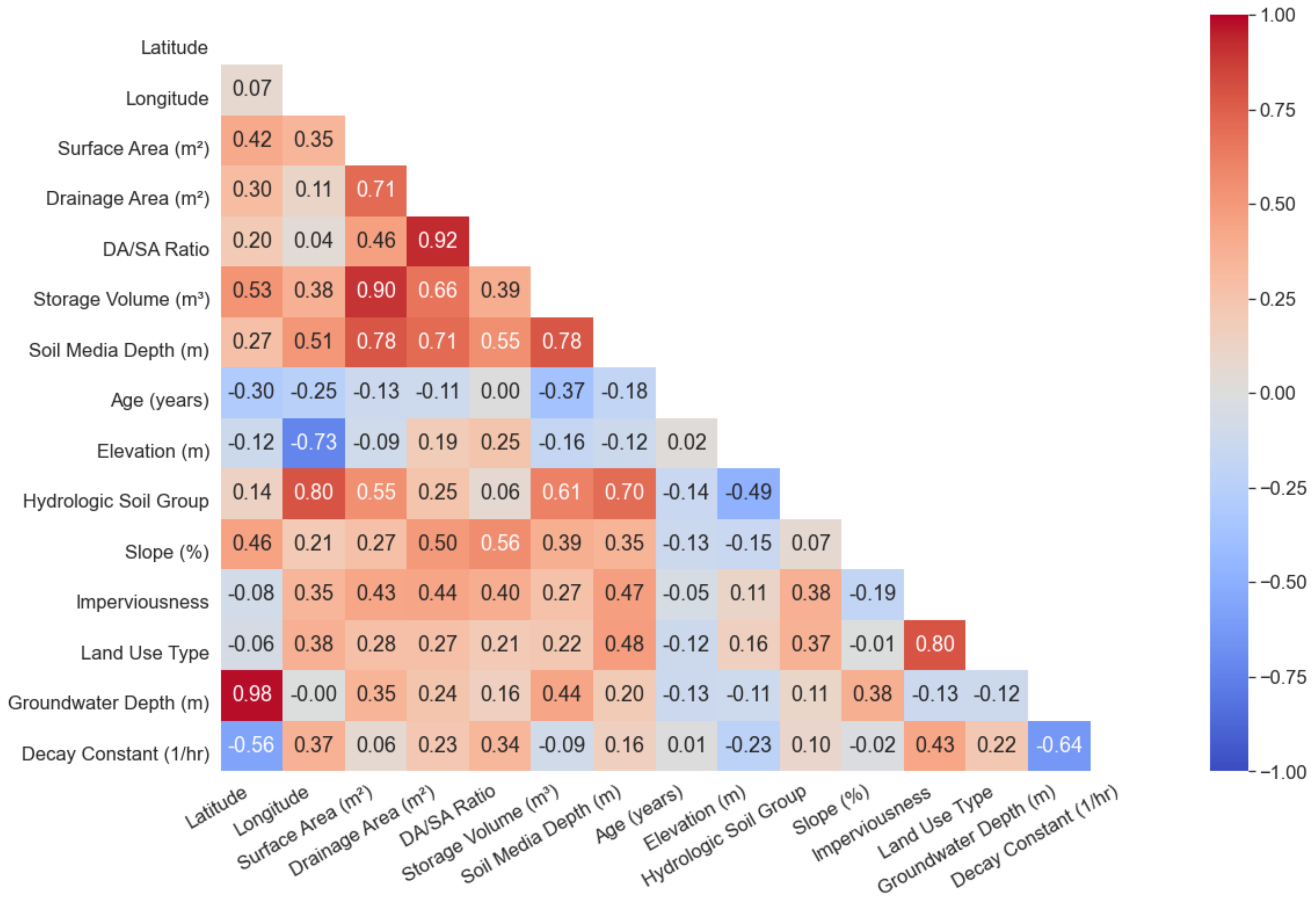}
  \caption{Spearman’s rank order correlation coefficients for the decay constants, design features, and physiographic features.}
  \label{fig:correlation}
\end{figure*}

\subsection{Correlation analysis}
\label{Results:correlation}
Spearman’s rank correlation coefficients between the GI design features and the decay constants ranged from 0.01 (site age) to 0.34 (DA/SA ratio) (Fig. \ref{fig:correlation}). The decay constants were most correlated with the DA/SA ratio (0.34) and drainage area (0.23). Drainage area and DA/SA ratio were highly correlated with each other (0.92); therefore, we focus analysis on the DA/SA ratio. The sites with the largest DA/SA ratios had the smallest magnitude decay constants (i.e., drained the slowest). Soil media depth, storage volume, surface area, and age had limited impact on the decay constants (0.16, \textminus0.09, 0.06, and 0.01, respectively).

The correlation coefficients between the physiographic features and the decay constants ranged from \textminus0.02 (slope) to \textminus0.64 (groundwater depth) (Fig. \ref{fig:correlation}). The decay constants were most correlated with groundwater depth (\textminus0.64), latitude (\textminus0.56), imperviousness (0.43), and longitude (0.37). The closer groundwater was to the surface, the slower the site drained (i.e., the smaller the decay constant’s magnitude). Groundwater is also highly correlated with latitude (0.98), which explains the correlation between latitude and the decay constants. Longitude, however, is not correlated with groundwater but still has a positive correlation with the decay constants. The decay constants’ magnitude decreases for sites further away from the western border towards central Detroit, where the smallest magnitude decay constants are, increasing again towards the eastern border. In terms of imperviousness, the greater the imperviousness, the smaller the decay constant’s magnitude. This was not always the case, however. For example, S1 and S12 are 53 and 52\% impervious and their mean $\alpha$'s are \textminus0.040 and \textminus0.047 hr$^{-1}$, respectively, while S9 is 92\% imperviousness with a mean $\alpha$ of \textminus0.102 hr$^{-1}$. The remaining physiographic features are either highly correlated with the explanatory variables discussed above (elevation and longitude: \textminus0.73; land use type and imperviousness: 0.80) or are minimally correlated with the decay constants (hydrologic soil group: 0.10; slope: \textminus0.02).

\begin{figure*}[!htb]
\centering
  \includegraphics[width=\textwidth]{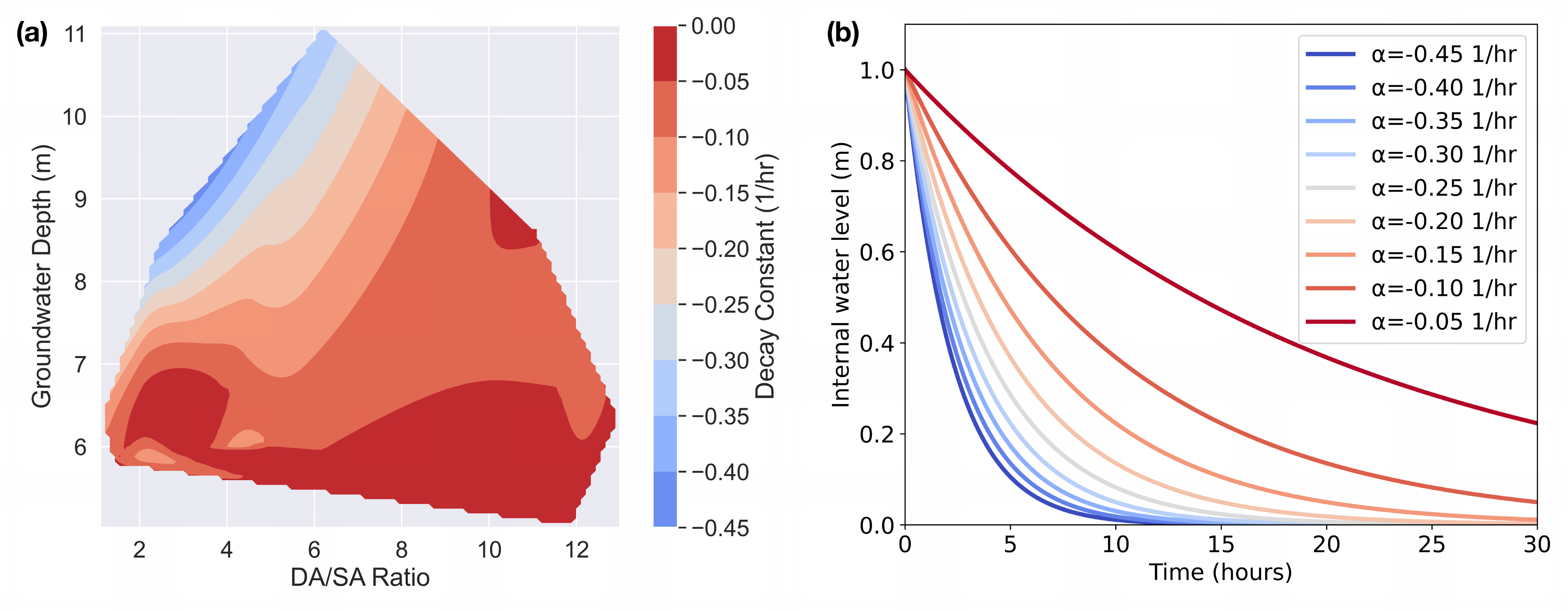}
  \caption{(a) A surface fit of the calculated decay constants (hr$^{-1}$) based on groundwater depth (m) (y-axis) and DA/SA ratio (x-axis). (b) The drawdown model curves for the range of decay constants found in (a). Blue indicates faster drawdown rates while red indicates slower rates.}
  \label{fig:gndwterVSdasa}
\end{figure*}

The relationship between the decay constant and its most correlated design feature, DA/SA ratio, and physiographic feature, groundwater depth, was explored further. We show groundwater depth versus DA/SA ratio for estimated decay constants in Fig. \ref{fig:gndwterVSdasa}a. Given that decay constants were retrieved for individual sites and individual storms, the figure reflects averaged surface fit across all the observations. The shape of Fig. \ref{fig:gndwterVSdasa}a is bounded by the observations made by the sensor network and was not extrapolated beyond those bounds. The colored contours indicate the expected decay constant based on the combination of groundwater depth and DA/SA ratio. The red contours indicate slower drawdown while the blue/grey contours indicate faster drawdown. To frame the interpretation of the figure, the corresponding drawdown rates are also color coded in (Fig. \ref{fig:gndwterVSdasa}b). 

In our study, decay constants with magnitudes $\geq$ \textminus0.20 hr$^{-1}$ result in the drainage of one meter of water in under 24 hours (Fig. \ref{fig:gndwterVSdasa}b). Fig. \ref{fig:gndwterVSdasa}a shows there are various combinations of groundwater depth and DA/SA ratio that achieve this performance metric. On one end of the spectrum, groundwater can be as shallow as 7.5 m if it has a small DA/SA ratio of 1--2. On the other end of the spectrum, groundwater must be at least 10 m deep with a DA/SA ratio no larger than 8. Furthermore, if the groundwater table is < 7.5 m, a slower drawdown rate is observed regardless of the DA/SA ratio (bottom edge of Fig. \ref{fig:gndwterVSdasa}a). Similarly, when the DA/SA ratio is >8, the drawdown rate is slow regardless of the groundwater depth (right edge Fig. \ref{fig:gndwterVSdasa}a).

\section{Discussion}
\subsection{GI drawdown dynamics}
The data toolchain introduced in this paper provides an automated way to analyze high resolution hydrologic data, such as water levels in GI. This is enabled by the storm segmentation methodology, which automatically extracts and analyzes data from individual storms. As sensor networks scale, manual data analysis will become infeasible, demanding that we discover means by which to automatically extract relevant data for analysis or training of machine learning algorithms becomes infeasible. As demonstrated here, the approach automatically identified storm events and subsequently analyzed them to train models for the decay constants. The application of a peak-find algorithm to extract events from other types of data (flows, rainfall, soil moisture, etc) should be explored in future studies.

The water levels from the 14 sensors indicate the GI are generally performing as designed, despite record rainfall. The GI met and exceeded the requirement specified by Detroit’s GI design manual that ponding time should not exceed 24 hours.\cite{water_and_sewerage_department_stormwater_2022} Below the ground surface, the performance varied by site and storm. To completely drain 1 m of water in 24 hours a GI must have a decay constant $\geq$ \textminus0.2 hr$^{-1}$ (Fig. \ref{fig:gndwterVSdasa}b). Only 2 of the 14 gardens had an average decay constant above this threshold. Therefore, most sites have restricted storage capacity when they experience consecutive storms.

Fitting a drawdown model for each storm and each site resulted in variability across decay constant estimates. Statistical uncertainty is inherent in a study of this scale, and may manifest across measurements, deployment consistency, and model assumptions. Some variability in the decay constants was likely due in part to the spatial and temporal variation in rainfall.\cite{cristiano_spatial_2017} The decay constants may also have been impacted by changes in GI conditions such as the swelling and shrinking of the soil media following wet and dry periods, and the creation of preferential flow paths after extended dry periods.\cite{hatt_hydraulic_2007} 

Naturally, a highly granular and continuous sensor dataset can be expected to reveal dynamics and nonlinearities that are not apparent in single measurements or short-term experimental campaigns. We contend that the use of the decay constant poses a first step in the analysis of this large dataset and provides an initial balance by enabling a metric for cross-site comparisons without compressing large amounts of sensor data into an over simplistic summary that ignores dynamics entirely. Future studies could explore the nuanced variabilities dynamics more explicitly.

Cross-site comparisons of water level dynamics revealed patterns driven by site design and physiographic features. It is difficult to directly attribute the variation seen between sites to the variations in these features due to the complexity of the physical processes that govern GI drainage dynamics. The correlation analysis found broadly, however, that GI with DA/SA ratios smaller than 8 have faster drawdown rates. Therefore, when designing GI, the size of the garden in relation to the size of the drainage area is critically important. These results align with Davis (2007),\cite{davis_field_2007} which found that a large cell media volume to drainage area ratio and drainage configurations were the two most dominant factors that improved GI performance. 

Across the broader landscape, GI drawdown dynamics were highly correlated with two physiographic features: groundwater depth and longitude. Faster drawdown rates were correlated with a deeper groundwater table and locations on the outskirts of Detroit. This illustrates the importance of evaluating groundwater levels when planning urban GI installations, especially since many urban areas have shallow groundwater tables,\cite{small_global_2003} including Detroit.\cite{teimoori_modeling_2021} The correlation with longitude may be explained by prolonged soil compaction from development in central Detroit.\cite{howard_evaluation_2016} 

Some physiographic features had low correlation with the decay constants. Detroit is relatively flat, which may explain the low correlation with elevation and slope. The low correlation between the decay constants and the hydrologic soil group of the surrounding soil is more difficult to posit. Our physiographic input data were limited to public datasets, whose accuracy is driven by factors outside of the control of this study. The low spatial resolution of publicly available raster datasets may oversimplify the physiographic features at a GI site. In the future, site surveys may provide better data for analyzing these physiographic features interaction with the decay constants.

Our results have several implications for the future of stormwater management. Considering the broader urban drainage landscape and the potential impact of physiographic features on GI drawdown rates, measurements should become a core component of how managers choose to invest in GI. For example, measuring the drawdown rate, groundwater depth, and/or soil compaction at a site before installation could reduce the risk of installing GI in locations that will have impeded drainage regardless of how well they are engineered. Beyond single sites, an investment into an entire measurement network may help support a more targeted and data-driven approach to GI placement, planning, and maintenance. The application of this methodology could result in empirical design guidance, such as an empirical “heatmap”, as shown in Fig. \ref{fig:gndwterVSdasa}a. Such illustrations could serve as a field-validated guide for managers who want to push the performance of their infrastructure without focusing all of their limited resources into one particular design or locale. Naturally, this would require the collection and analysis of more data, but the increasing reliability of technology and automation afforded by some of the tools in this paper may reduce the barrier to adoption.

One potential limitation of this work is the duration of our study period. Over longer periods of time we would expect to see fluctuations in the decay constants due to seasonal conditions (e.g., the rate of evapotranspiration falling during colder months\cite{spraakman_how_2021}) and due to longer-term trends (e.g., deterioration of the GI’s drainage capacity due to clogging\cite{taguchi_it_2019}). In future work, how the decay constants vary over time should be investigated to determine these seasonal and long-term changes. The reliability of the sensors should enable long-term data collection with reduced measurement overhead.

\subsection{Beyond site-level drawdown dynamics}
This study used the high temporal and spatial resolution dataset produced by a sensor network to provide a first order analysis of the variability in GI drawdown dynamics, but the sensor network could also be used for a variety of other purposes. Large GI sensor networks have potential for use in long-term GI monitoring. These data can used to develop a deeper understanding of how GI installations fit into the larger urban drainage network, but this may also require the application of expanded tools for data analysis. Given the accessibility to and availability of modern Machine Learning libraries, the data collected by these networks could be used to inform predictive tools and interactive design guides. The sensor data can also be used to iterate on site design or inform maintenance schedules. Measurements showing when drainage slows over time could indicate that the GI soil media is clogged and should be replaced. A science-based method to validate such scenarios should be investigated. These data may also be used for community education and engagement by communicating to residents and community groups how and where GI may be expected to work well.

\section{Conclusion}
This study introduces a wireless, real-time sensor for measuring GI drawdown. Networked together across Detroit, these sensors provide high temporal and spatial resolution data for analyzing city-scale urban drainage conditions. To isolate individual storms in this large dataset, we designed an automated storm segmentation methodology based on peak finding. To our knowledge, this study is the first to monitor GI at this scale and combine it with a data-driven workflow to reveal explanatory features of drawdown dynamics. In Detroit, the groundwater table, imperviousness, longitude, and DA/SA ratio are the most important features impacting drawdown rates. To confirm this finding for other regions, high resolution and long-term GI monitoring is necessary.

\section*{Author Contributions}
\; \; \textbf{Brooke E. Mason:} Conceptualization; Methodology, Software, Validation, Data Curation, Formal Analysis, Investigation, Writing – Original Draft, Writing – Reviewing and Editing, Visualization, Supervision

\textbf{Jacquelyn Schmidt:} Conceptualization, Methodology, Software, Data Curation, Writing – Original Draft, Writing – Reviewing and Editing, Visualization

\textbf{Branko Kerkez:} Conceptualization, Methodology, Resources, Writing – Reviewing and Editing

\section*{Conflicts of interest}
There are no conflicts to declare.

\section*{Acknowledgements}
We would like to thank our collaborators Erma Leaphart, Elayne Elliott, and Cyndi Ross with the Detroit Sierra Club. We would like to thank all the site owners for allowing us to install sensors at their homes, churches, and schools. We would like to thank Angela Hojnacki, Ian Thompson, and Kevin Kaya for installing the sensor network. We would like to thank Lance Kruse for his expertise in statistics. Finally, we would like to thank our Project Manager, Kate Kusiak Galvin. This work was funded by the U.S. National Science Foundation (Award Numbers: 1737432 and 1750744).



\balance


\bibliography{rsc} 
\bibliographystyle{rsc} 

\end{document}